\documentclass[a4paper, 12pt]{article}
\begin{document}

\title{Ion Trap Proposal for Quantum Search}
\author{Joonwoo Bae{\footnote{Email address: jwbae@newton.hanyang.ac.kr}}, Younghun Kwon{\footnote{ Email address: yyhkwon@hanyang.ac.kr}} \\Department of Physics, Hanyang University, \\Ansan, Kyunggi-Do, 425-791, South Korea}

\maketitle
\vspace{-.4in}
\begin{abstract}
In this letter, we show that the laser Hamiltonian can perform the quantum search.
We also show that the process of quantum search is a resonance between the initial state and the target state, which implies that Nature already has a quantum search system to use a transition of energy. In addition, we provide the particular scheme to implement the quantum search algorithm based on a trapped ion. 
\end{abstract}

Quantum computation has been in the spotlight, supplying the solution for the problems which are intractable in the context of classical physics. The quantum factorization algorithm and the quantum search algorithm are the good examples.[1] In particular, the quantum search algorithm provides the quadratic speedup in solving a search problem. Here, the search problem is to find the target of the unstructured $N$ itmes. Grover's fast quantum search algorithm is composed of discrete-time operations(e.g., Walsh-Hadamard). When we use Grover's algorithm, we need $O(\sqrt{N})$ iterations of the Grover operator. There is the analog quantum search algorithm which is based on the Hamiltonian evolution. Farhi and Gutmann proposed the quantum search Hamiltonian.[5] Moreover, Fenner provided another quantum search Hamiltonian.[6] Recently, the generalized quantum search Hamiltonian was proposed.[7] \\
 In this letter, we show that a laser Hamiltonian can perform quantum search by resonance between the initial state and the target state. The idea to apply the laser Hamiltonian to quantum search comes from the observation that the generalized quantum search Hamiltonian is apparently similar to the laser one. The difference is that the quantum search Hamiltonian is time-independent but the laser Hamiltonian time-dependent. In order to develop our proposition, we first introduce the generalized quantum search Hamiltonian. Next, we briefly review the laser Hamiltonian and show that the laser Hamiltonian can perform the quantum search. Also we point out that the resonance is the key factor in analog quantum search. Furthermore, we suggest the scheme to implement the quantum search algorithm using a trapped ion. \\
 In order to apply quantum mechanics to a search problem, we give a bijection from the $N$ items to the eigenstates of the $N$-dimensional Hilbert space. The search problem is then to find the target state in the $N$ eigenstates. Grover algorithm is the method that offers the quadratic speedup by quantum mechanics. It is composed of three procedures - initialization, Grover iterations, and readout.[1-3] The first procedure is to prepare an initial state.[2] Usually, the initial state is uniformly superposed with the $N$ states since there is no information about the target. The next procedure is to apply Grover iterations and the final procedure to measure the resulting state. There are two stages in the Grover iteration. The first stage is to flip the target, through the oracle function $f(k) = \delta_{k, \alpha}$, where $k$ stands for any item and $\alpha$ the target item. The next stage is to invert the current state about the average of all states.[3] After single Grover iteration is performed, the amplitude of a target state is boosted by the amount of $1/O(\sqrt{N})$. Thus an initial state comes to a target state by $O(\sqrt{N})$ Grover iterations. It is the oracle function that is the key to the speedup in Grover algorithm, since it only identifies the target state. Zalka proved that the running time of Grover algorithm is optimal.[4] Another quantum search algorithm based on Hamiltonian evolution, which is composed of continuous-time operations(analog computation), also provides the quadratic speedup. \\
 Any quantum search Hamiltonians[5-6] are specific cases of the generalized quantum search Hamiltonian $H_{g} = E(|\alpha \rangle \langle \alpha|+|\psi \rangle \langle \psi|)+ \epsilon (e^{i\phi}|\alpha \rangle \langle \psi|+e^{-i\phi}|\psi \rangle \langle \beta|)$[7], where $| \alpha \rangle$ is the target state, $| \psi \rangle $ the initial state, $E$ and $\epsilon$ are arbitrary constants in unit of energy, and $\phi$ is a phase. When $E$, $\epsilon$, and $\phi$ are arbitrarily chosen and the initial state is the uniform superposition of $N$ states, the Hamiltonian finds the target state within $O(\sqrt{N})$ times with probability $1-O(1/N^{2})$. Furthermore, the target state is obtained with probability one, only when the phase is fixed as $\phi = n \pi$. \\
 Let us consider a laser Hamiltonian $H_{laser} = E_{a}|a \rangle \langle a| + E_{b}|b \rangle \langle b| + \gamma ( e^{iwt}|a \rangle \langle b|+e^{-iwt}|b \rangle \langle a|)$, where $E_{a}$ and $E_{b}$ are energy eigenvalues.[8,9] Note that, a laser Hamiltonian describes the transition of the state from $| a \rangle$ to $| b \rangle$ by the resonance between the two states, only when the frequency satisfies $w = \frac{E_{b} - E_{a}}{\hbar}$(resonance condition). Observing $H_{g}$ and $H_{laser}$, we expect that the laser Hamiltonian would solve a search problem by overcoming the difficulty that $H_{g}$ is time-independent but $H_{laser}$ is not. We decompose the initial state in the quantum search Hamiltonian by the target state and its orthogonal complement, that is, $|\psi \rangle = x |\alpha \rangle + \sqrt{1-x^{2}}|\beta \rangle$, where $x = \langle \alpha | \psi \rangle $. Since the quantum search Hamiltonian is time-independent and the laser Hamiltonian time-dependent, we expand the phase factor $\phi$ in time as a linear function of time, $\phi = wt + \varphi$. Then, we obtain the following Hamiltonian $H_{ls}$ :

\begin{displaymath} 
H_{ls} = E_{\alpha}|\alpha \rangle \langle \alpha| + E_{\beta}|\beta \rangle \langle \beta| + \gamma (e^{i(wt+\varphi)}|\alpha \rangle \langle \beta |+e^{-i(wt+\varphi)}|\beta \rangle \langle \alpha|)
\end{displaymath}

, where $E_{\alpha} = Ex + \epsilon cos \phi$, $E_{\beta} = E(1-x^{2})$, $\gamma = \sqrt{1-x^{2}} \sqrt{(Ex+\epsilon cos \phi)^{2} + (\epsilon sin \phi)^{2}  }$, and $\varphi = sin^{-1} \frac{\epsilon sin \phi}{\sqrt{(Ex+\epsilon cos \phi)^{2} + (\epsilon sin \phi)^{2}  }}$. These conditions mean that the initial state is the superposition of $|\alpha \rangle$ and $|\beta \rangle$. It is easy to show that the Hamiltonian $H_{ls}$ performs the quantum search by the transition of states from $|\beta \rangle$ to $|\alpha \rangle$, fixed by the frequency $w = \frac{2x}{\hbar} (-Ex-\epsilon cos \phi)$(resonance condition). Since the frequency $w = \frac{2x}{\hbar} (-Ex-\epsilon cos \phi)$ is positive, it satisfies $-1 \leq cos \phi < -\frac{Ex}{\epsilon}$. This inequality implies that $\epsilon > Ex$ is necessary. Thus we can conform that the Hamiltonian which is induced from the generalized quantum search Hamiltonian performs the quantum search. In other words, a laser Hamiltonian is
the linearly time-dependent version of the generalized quantum search Hamiltonian. \\
 Here, we propose a scheme to implement the quantum search algorithm based on the evolution of the Hamiltonian $H_{ls}$, using a trapped ion. Ion trap quantum computation makes use of the resoance phenomenon for bit-operations.[10,11] With the usual conventions of an ion trap quantum computer, the computational basis $|1 \rangle$ and $|0 \rangle$ are internal levels of a trapped ion, a ground state and an excited state, respectively. The Hamiltonian for the static state of an ion is $H_{ion} = E_{0}|0\rangle \langle0| + E_{1}|1\rangle \langle1|$. By applying $V-pulse$ of the frequency $w = \frac{E_0 - E_1}{\hbar}$(resonance condition) to the ion, the time-dependent Hamiltonian $U(t) = \gamma (e^{iwt}|0\rangle \langle 1|+e^{-iwt}|1 \rangle \langle 0|)$ is induced. Then the full Hamiltonian becomes $H(t) = H_{ion} + U(t)$. Thus we can flip a bit $|0\rangle$ to $|1\rangle$ with the $V-pulse$(single qubit operation). The strategy to implement the quantum search algorithm with a trapped ion is to regard the excited state $|0\rangle$ as a superposition of adequate $N-1$ orthogonal state excluding the ground state $| 1 \rangle$, i.e., $|0\rangle = \frac{1}{\sqrt{N-1}} \sum_{i=1}^{N-1} |k_{i} \rangle$. (Here we regard the target $|\alpha \rangle$ as  $| 1 \rangle$, that is $| \alpha \rangle = | 1 \rangle$)
At equilibrium(before starting computation), the Hamiltonian for the static state of an ion is $H_{s}  = (E(1+\frac{1}{N})-2\epsilon \frac{1}{\sqrt{N}})|\alpha \rangle \langle \alpha| + E(1-\frac{1}{N})|\beta \rangle \langle \beta|$. To find the state $|\alpha \rangle$, we apply the $V-pulse$ of the frequency $w = \frac{2}{N \hbar} (\sqrt{N}\epsilon -E)$ with $\epsilon > \frac{E}{\sqrt{N}}$ to the ion. The time-dependent term $\sqrt{1-\frac{1}{\sqrt{N}}}(\frac{E}{\sqrt{N}}-\epsilon)(e^{iwt}|\alpha \rangle \langle \beta|+e^{-iwt}|\beta \rangle \langle \alpha|)$ is induced. Thus the full Hamiltonian is :

\begin{displaymath}
H_{s}  = (E(1+\frac{1}{N})-2\epsilon \frac{1}{\sqrt{N}})|\alpha \rangle \langle \alpha| + E(1-\frac{1}{N})|\beta \rangle \langle \beta| + \sqrt{1-\frac{1}{\sqrt{N}}}(\frac{E}{\sqrt{N}}-\epsilon)(e^{iwt}|\alpha \rangle \langle \beta|+e^{-iwt}|\beta \rangle \langle \alpha|)
\end{displaymath}
, which is the specific case of $H_{ls}$ by fixing the phase $\phi = (2n+1)\pi$ and setting the initial state $|\psi \rangle = \frac{1}{\sqrt{N}}(|\alpha \rangle + |\beta \rangle) = \frac{1}{\sqrt{N}} \sum_{j=0}^{N-1} |k_{j}\rangle$. We obtain the target state with probability one, by measuring the resulting state only after $O(\sqrt{N}/E)$ evolution times. Note that the frequency is the function of the number of state $N$ and two values $E$ and $\epsilon$ in unit of energy, and therefore can be controlled by the values of $E$, $\epsilon$, and $\phi$. \\
 The fact that a quantum search Hamiltonian is the laser Hamiltonian that describes the transition of energy from an initial to a target state, implies that only a target state should be resonated with a current state. In other words, it means that the key in quantum search hamiltonian is resonance. Since it is the oracle that plays the role of distinguishing a target state from the others, we can claim that the oracle in the time-dependent quantum search Hamiltonian $H_{ls}$ is implicitly represented as $f(k) = \delta_{k,\alpha}$. Remind that the oracle function is used in Gover operator.
 By noting that Grover algorithm is obtained by discretizing Schrodinger equation in time,[12] we can find out that the role of the oracle in all quantum search algorithms makes a current state  resonated with a target state. That is, the role of oracle is to single out the target in the database by resonance. Amplitude amplification by Grover iterations is the discrete-time version of the transition of energy. \\
 In this letter, we have shown that the laser Hamiltonian can perform the quantum search and illustrated that the feature of the quadratic speedup in quantum search is the resonance between the initial state and the target state. Thus, the oracle in a quantum search Hamiltonian has the representation $f(k) = \delta_{k, \alpha}$ implicitly.
We also suggested a specific scheme to implement a quantum search algorithm. Note that the phase factor in $H_{ls}$ can be controlled, since it is the function of $N$, $E$, and $\epsilon$. We can expect that this result should provide a new approach to developing a quantum search device. \\
Ion trap quantum computation that was proposed in this paper is only an example. Any two-level systems that describe the transition of energy by resonance have the chance to implement the quantum search algorithm. Therefore, we are expecting a more practical design to implement the algorithm for a current device. Otherwise appropriate equipment must be invented for the implementation. The best way is to find an intrinsic quantum search system. 
\\

\section*{Acknowledgement}
J. Bae is supported in part by the Fellowship of Hanyang University and Y. Kwon is supported in part by the Fund of Hanyang University.

\end{document}